# Defining the sample designs for small area estimation

Falorsi Piero Demetrio, Stefano Falorsi, Vincenzo Nardelli and Paolo Righi[1]

**Abstract.** The paper delineates a proper statistical setting for defining the sampling design for a small area estimation problem. This problem is often treated only via indirect estimation using the values of the variable of interest also from different areas or different times, thus increasing the sample size. However, let us suppose the domain indicator variables are available for each unit in the population. In that case, we can define a sampling design that fixes the small-area samples sizes, thereby significantly increasing the accuracy of the small area estimates. We can apply this sampling strategy to business surveys, where the domain indicator variables are available in the business register and even household surveys, where we can have the domain indicator variables for the geographical domains. The two primary steps of sampling processes that characterize the sampling strategy we propose are the optimization step and the sampling selection. The optimization step defines and solves a non-standard allocation problem that finds the minimum-costs solution that guarantees that the accuracy of the model-based estimator is lower than pre-specified thresholds for each domain of interest. To allow the re-use of the methodology in other contexts, we developed an R package with a set of functions to solve the optimization problems. The sampling selection is carried out through incomplete stratification techniques. These techniques control the sample sizes only at the domain level, avoiding the problems caused by an overly detailed stratification by cross-classifying the domains and other design variables. This stratification is often unsuitable. Indeed, it requires the selection of at least several sampling units as large as the product of the number of categories of the stratification variables. Subsequently, we present empirical shreds of evidence on a simulated data-set, confirming both the efficacy of the proposed methodology and the algorithmic solution's convergence.

[1] Piero Demetrio Falorsi, International consultant; piero.falorsi@gmail.com
Stefano Falorsi, Senior researcher, Istat; stfalors@istat.it
Vincenzo Nardelli, researcher, Università degli Studi di Milano-Bicocca; v.nardelli2@campus.unimib.it
Paolo Righi, Senior researcher, Istat; parighi@istat.it.

**Statement of significance.** When dealing with a small area estimation problem, relying solely on the indirect estimate is usual. Indirect estimation involves combining values of the variable of interest from multiple areas or timeframes, thus increasing the sample size. Let us suppose that the domain indicator variables are accessible for each unit in the population. In that case, we may create a sampling strategy that defines the small-area sample sizes and dramatically improves the accuracy of the estimates. We can adopt this strategy in business surveys, where we can find domain indicator variables in the business registry, and even in household surveys, where domain indicator variables are available for geographical domains. Two key steps characterize the suggested sampling technique: optimization and sampling selection. The optimization solves an allocation problem by discovering the minimum-cost option while ensuring that the model-based estimator's accuracy is below pre-determined thresholds. We created an R package with various functions to solve the optimization challenges. The sample selection is made using marginal-stratification techniques, which avoid the issues of excessively detailed stratification. Finally, we present the findings of a simulation study, which demonstrate the efficacy of the proposed methodology.

# 1 Introduction

This paper deals with the problem of defining the sampling design in the case in which we estimate the characteristics of interest with Small Areas Estimation (SAE) techniques (Rao, 2003) developed under a model-based approach to inference (Chambers and Clark, 2015).

We denote a sub-population as a small area (or small domain) if a direct estimator (for instance, the Horvitz-Thompson estimator, calibration estimator, etc.), essentially based on the sampling design (Särndal et al., 1992), produces unreliable estimates. Moreover, let us suppose that we do not observe a domain in the sample. In that case, we cannot compute the direct estimate since the applicability of the estimator depends on the presence of units in the domain of interest. (Lehtonen et al. l. 2003; Lehtonen and Veijanen, 2009; Rao, 2015).

According to a model-based approach, we often treat the SAE problem via indirect estimation using the values of the variable of interest $y$ also from different areas or different times, thus increasing the sample size (Rao, 2003). However, if the domain indicator variables are available for each unit in the population, there are opportunities to be exploited at the survey design stage. This condition often characterizes some surveys, where the domain indicator variables are available in the business register and even in the household surveys where we can have the domain indicator variables for the geographical domains. Singh, Gambino and Mantel (1994) noted a need to develop an overall strategy that deals with the SAE problem, involving planning sample design and estimation aspects. A consolidated approach (see Chambers and Clarck, 2012; Valliant, 2000 Dorfmann and Valliant, 2005) for constructing this strategy recommends using the models to predict values for the non-sampled units of the population. However, in such model-based inference, it is crucial to guard against problems arising from the model failure. Selecting a random balanced sample or weighted balanced sample provides bias protection in the relevant case of estimating means or totals. Pfeffermann and Ben-Hur (2019) contribute to the same topic of joint use of random sampling and model-based inference proposing a hybrid inferential approach to estimating the design-based mean squared error (DMSE) of model-dependent small area predictors.

The analysis of the model-based estimation bias due to model failure and the sampling process is not in the scope of the paper. Here, we are interested in studying how the sample units in each domain of interest enable a gain in efficiency of the indirect estimates (Pfeffermann, 2015). In this framework, we could treat the small domains as planned domains, for which the sampling design defines the sample size. Indeed, having sample units in each domain would allow us to estimate a specific small area random effect for each domain of interest. This condition provides more accurate indirect estimates for all domains than the situation in which we do not observe some domains in the sample. In the latter case, in fact, it is impossible to estimate the area effect for the unobserved domains. Therefore, the indirect estimate is potentially less accurate as it is based only on the synthetic estimator. Consequently, obtaining an accurate estimate of the random area effect for each domain of interest produces a more robust estimator that can potentially prevent failures of the small area model at the basis of estimation.

Coming back now to the problem of allocation, which is the main object of this work, we remember that Marker (2001) first contributed to this topic by proposing a cross-classification sampling design where strata are identified by cross-classifying the domains and other design variables. The author notes that stratifying and oversampling small domains makes it possible to significantly improve the accuracy of direct estimates for these domains while incurring a minimal loss in accuracy for national estimates. However, in many practical situations, this design is unsuitable since it requires the selection of at least several sampling units as large as the product of the number of categories of the stratification variables (Cochran 197, page 124).

Falorsi and Righi (2008) overcome the weakness of the cross-classified stratified design, introducing the use of the incomplete stratified design (Falorsi and Righi, 2015). The incomplete stratification takes the cross-classified stratified design into account. The incomplete stratification also controls the sample sizes within the marginal strata corresponding to the domain partitions, while the sample size in the cross-classified strata is a random number. The proposed sampling strategy uses a balanced sampling selection technique (Deville and Tillé, 2004) and a GREG-type estimation (Lehtonen et al., 2003; Lehtonen and Veijanen, 2009). The sample allocation in the strategy specifies the sample sizes in the marginal strata that guarantee that the sampling errors of the domain estimators are lower than pre-specified thresholds. Furthermore, Falorsi and Righi (2008) introduce the idea to extend the strategy to synthetic estimators and model-based estimation.

Burgard et al. (2014) and Friedrich et al. (2018) focus on a topic related to optimal SAE sampling. In both papers, the authors assume the necessity to provide figures on domain levels and a variety of subclasses. The use of cross-classified design leads to fine stratifications of the population. Optimizing the accuracy of stratified random samples requires incorporating a vast amount of strata on various levels of aggregation. Taking into account several variables of interest for the optimization yields a multivariate optimal sampling allocation. In particular, Burgard et al. (2014) study the effect of some sampling designs on SAE for business data and suggest considering the box-constraint optimal allocation proposed by Gabler et al. (2012). Keto et al. (2018) recommend a sampling allocation based on multi-objective optimization using a small-area model and estimation method introduced by Falorsi and Righi (2008).

This paper further develops this setting. We study the problem of defining the minimum cost sampling solution for model-based inference. The sample allocation specifies the sample sizes in the domains that guarantee that the MSE of the indirect model-based estimator is lower than pre-specified thresholds. Moreover, to carry out the experimental results presented in the paper and allow for the re-use of the methodology in other contexts, we developed the R package *saeall* [3] with a set of functions to solve the optimization problems.

We organized the paper as follows. Section 2 presents the statistical setting, describing the parameter of interest and the sampling. Section 3 introduces the sampling optimization problem under the basic random-mean model. Section 4 proposes computational simplifications that dramatically reduce the number of parameters to estimate the optimization problem. Sections 5 and 6 suggest the extension to the two-stage sampling and for dealing with the uncertainty of domain membership variables. Finally, Section 7 concludes the theoretical part of the paper by considering general small area unit-level models. We propose a fixed-point algorithm for dealing with optimization problems for these cases. Section 8 presents empirical evidence on a simulated data set, confirming both the efficacy of the proposed and the algorithmic solution's convergence.

## 2  Statistical setting

*Target parameters*

Let $U$ be a population of size $N$, and let $y_{(v)}$ ($v = 1, \ldots, V$) be a target variable. Let $y_{(v)i}$ ($v = 1, \ldots, V; i = 1, \ldots, N$) indicate the value of $y_{(v)}$ in the $i-th$ unit of $U$. Let $U_d$ ($d = 1, \ldots, D$) be a particular sub-population of $U$ of size $N_d$ and let $\lambda_{di}$ ($d = 1, \ldots, D; i = 1, \ldots, N_d$) be the unit *domain* membership variable, where $\lambda_{di} = 1$ if $i \in U_d$ and $\lambda_{di} = 0$.

The different domains can overlap, so it is possible that $\lambda_{di}\lambda_{d'i} = 1$ for $d \neq d'$. The domains may define alternative partitions of $U$. For example, regions can define a specific partition of domains in business surveys. Similarly, the classes of the Nace -activity codes (Eurostat, 2008) individuate an alternative partition. Let

---
[3] https://github.com/vincnardelli/saeall

$$(2.1) \quad Y_{(v)d} = \sum_{i \in U} y_{(v)i} \, \lambda_{di} \quad (v = 1, \ldots, V; d = 1, \ldots, D)$$

be the target parameters of interest for the domain $U_d$. Expression 2.1 identifies a multivariate-multi-domain situation. Indeed, there are $V$(variables) $\times$ $D$(domains) target parameters.

*Sampling*

A sample $S$ of size $n$ is selected without replacement from $U$ with inclusion probabilities $\boldsymbol{\pi} = (\pi_1, \ldots, \pi_i, \ldots, \pi_N)'$.

We suppose that the domain indicator variables $\boldsymbol{\lambda}_i = (\lambda_{1i}, \ldots, \lambda_{di}, \ldots, \lambda_{Di})'$ are available in the sampling frame. We furthermore assume selecting the sample $S$ while respecting the following balancing equations:

$$(2.2) \quad \sum_{i \in S} \frac{\boldsymbol{b}_i}{\pi_i} \cong \sum_{i \in U} \boldsymbol{b}_i$$

where $\boldsymbol{b}_i$ is a vector of $B$ auxiliary variables for the unit $i$. If we define the balancing variables in Expression 2.2 as

$$(2.3) \quad \boldsymbol{b}_i = \pi_i \boldsymbol{\lambda}_i,$$

then the sampling selection ensures planned sample sizes, $n_d$ $(d = 1, \ldots, D)$ for each domain Deville and Tillé (2004; p. 905 Section 7.3) have proven that (*i*) with the Specification 2.3 and (*ii*) if the vector of the expected sample sizes, given by

$$(2.4) \quad \boldsymbol{n} = (n_1, \ldots, n_d, \ldots, n_D) = \sum_{i \in U} \pi_i \, \boldsymbol{\lambda}_i,$$

includes only integer numbers, then a balanced sampling design always exists and the Equation 2.2 is exactly satisfied. Deville and Tillé (2004, pp. 895 and 905), Deville and Tillé (2005, p. 577) and Tillé (2006, p. 168) have shown that several customary sampling designs may be considered special cases of balanced sampling, by properly defining the vectors $\boldsymbol{\pi}$ and $\boldsymbol{\lambda}_i$ of Equation 2.2. Balanced samples may be drawn by means of the Cube method (Deville and Tillé,

2004). This enormously facilitates the sample selection. The Cube method satisfies 2.4 exactly when 2.4 holds and $\boldsymbol{n}$ is a vector of integers.

Finally, we note that a safe sampling strategy in small area estimation should ensure the sample is well spread on the space of auxiliary variables predictive of the target ones. Grafström and Tillé (2012) proposed the Local Cube (LC) method, which enables the selection of samples that are balanced on several auxiliary variables and at the same time are well spread for some variables, which can be geographical coordinates. Moreover, Grafström and Lündstrom (2013) have demonstrated that well-spread balanced samples in space are balanced on auxiliary variables even if the target parameters are nonlinear in the auxiliary variables.

*Model*

Consider for now a simple *random-mean* model (Model 15.11, Chambers and Clark, 2012)

(2.5) $y_{(v)i} \lambda_{di} = \mu_{(v)} + u_{(v)d} + e_{(v)i}$,

where $\mu_{(v)}$ is the mean of the population model, $u_{(v)d}$ indicates a random domain effect, and $e_{(v)i}$ a random noise. More complex models will be considered in Section 7. The random domain effects $\{u_{(v)d}; v = 1, \ldots, V; d = 1, \ldots, D\}$ are independently and identically distributed with model-expectation and variance $E_M(u_{(v)d}) = 0$ and $V_M(u_{(v)d}) = \sigma^2_{(v)u}$. The random individual effects $\{e_{(v)i}; v = 1, \ldots, V; i = 1, \ldots, N\}$ are also independently and identically distributed with $E_M(e_{(v)i}) = 0$ and $V_M(e_{(v)i}) = \sigma^2_{(v)}$.

Assuming that the variance components $\sigma^2_{(v)u}$ and $\sigma^2_{(v)}$, are known, the Best Linear Unbiased Predictor (BLUP) of $Y_{(v)d}$ is

(2.6) $\hat{Y}_{(v)d} = \sum_{i \in S} y_{(v)i} \lambda_{di} + \sum_{i \in U \setminus S} (\hat{\mu}_{(v)} \lambda_{di} + \hat{u}_{(v)d} \lambda_{di})$

where $\hat{\mu}_{(v)}$ is the Best Linear Unbiased Estimate (BLUE) estimate of $\mu_{(v)}$, and $\hat{u}_{(v)d}$ is the Best Linear Unbiased Predictor (BLUP) of $u_{(v)d}$. A standard result (Chambers and Clark, 2012, pg.170-171), states

$$(2.7) \quad E_M\left(\hat{Y}_{(v)d} - Y_{(v)d}\right)^2 = MSE_M(\hat{Y}_{(v)d}) = g1_{(v)d} + g2_{(v)d},$$

where

$$(2.8) \quad g1_{(v)d} = (N_d - n_d)^2 \frac{\sigma_{(v)}^2}{n_d} \gamma_d = (N_d - n_d)^2 \frac{\sigma_{(v)u}^2 \sigma_{(v)}^2}{n_d \sigma_{(v)u}^2 + \sigma_{(v)}^2},$$

$$(2.9) \quad g2_{(v)d} = (N_d - n_d)^2 \frac{\sigma_{(v)}^2}{n} (1 - \gamma_{(v)d})^2 \left(1 - n^{-1} \sum_{\kappa \in Par(d)} n_\kappa \gamma_\kappa \right),$$

where $n_d$ is the number of sample units that belong to the small domain $U_d$,

$$(2.10) \quad \gamma_{(v)d} = \frac{\sigma_{(v)u}^2}{\frac{\sigma_{(v)}^2}{n_d} + \sigma_{(v)u}^2},$$

and $Par(d)$ is the sub-set of the $D$ domains that include the sub-population $U_d$ and defines a partition of $U$, being $\kappa$ a generic domain of $Par(d)$. For instance, if $U_d$ is a given region, then $Par(d)$ is the set of the regions. We see that $g1_{(v)d}$ is of order $(N_d - n_d)^2 n_d^{-1}$, while $g2_{(v)d}$ is of order $(N_d - n_d)^2 n^{-1}$. Therefore, in most situations $g1_{(v)d}$ dominates $g2_{(v)d}$.

## 3 Sampling optimization problem under the random-mean model

Assume that the sampling selection ensures the planned sample sizes for each domain $d$ selecting the sample with balancing Equations 2.4.

Consider Expression 2.8 and assume that the sampling fractions $n_d/N_d$ are negligible and close to 0, as usual in most SAE problems. According to Rao (2015, 7.2.11, page. 137), with $n_d/N_d$ negligible, we can approximate expression 2.8 by substituting the squared difference

$(N_d - n_d)^2$ with the term $N_d^2$. We can express $g1_{(v)d}$ as a function of the inclusion probabilities $\boldsymbol{\pi}$:

$$(3.1) \quad g1_{(v)d} \cong N_d^2 \frac{\sigma_{(v)u}^2 \sigma_{(v)}^2}{n_d \sigma_{(v)u}^2 + \sigma_{(v)}^2} = N_d^2 \frac{\sigma_{(v)u}^2 \sigma_{(v)}^2}{(\sum_{i \in U} \pi_i \lambda_{di}) \sigma_{(v)u}^2 + \sigma_{(v)}^2}$$

The term $g2_{(v)d}$ is of minor order and it vanishes for large sample sizes $n$. The main sampling problem is defining the vector of inclusion probabilities $\boldsymbol{\pi}$ so as to have the expected minimum cost sampling design ensuring that the terms $g1_{(v)d}$ $(v = 1, \ldots, V; d = 1, \ldots, D)$ of $MSE_M(\hat{Y}_{(v)d})$ are lower than pre-fixed thresholds $g1_{(v)d}^*$ $(v = 1, \ldots, V; d = 1, \ldots, D)$.

The Optimization Problem (OP) may be defined as

$$(3.2) \quad \begin{cases} Min \left( \sum_{i \in U} C_i \pi_i \right) \\ N_d^2 \frac{\sigma_{(v)u}^2 \sigma_{(v)}^2}{(\sum_{i \in U} \pi_i \lambda_{di}) \sigma_{(v)u}^2 + \sigma_{(v)}^2} \leq g1_{(v)d}^* \text{ for } v = 1, \ldots, V; d = 1, \ldots, D \\ 0 < \pi_i \leq 1 \qquad\qquad\qquad\qquad\qquad \text{ for } i = 1, \ldots, N \end{cases}$$

where $C_i$ is the unit cost for surveying the unit $i$. In Problem 3.2, the variances $\sigma_{(v)u}^2$ and $\sigma_{(v)}^2$ are treated as known; in practice, they must be estimated. The main issue is to find an algorithmic solution to Problem 3.2 that represents a non-standard problem. In the context of inference based on the sampling design, we can define a similar problem, except that the constraints are different (Bethel, 1989; Chromy, 1987; Falorsi and Righi, 2015).

To solve Problem 3.2, we considered the iterative algorithm by Powell (1994) for nonlinearly constrained optimization calculations without using the derivatives. Each iteration forms linear approximations to the objective and constraint functions by interpolation at the vertices of a simplex and a trust region bound restricts each change to the variables. Thus, the algorithm calculates a new vector of variables, which may replace one of the current vertices, either to improve the simplex shape or because it is the best vector found so far. The algorithm is available in the NLopt library for non-linear optimization.[4] In the empirical experiment the R implementation available in the package nloptr [4] was used. Section 7 below presents a different fixed-point algorithm to solve problem 3.2.

## 4 Computational simplification

Problem 3.2 is computationally complex since there are $N$ unknown values. We can simplify the problem by $(i)$ partitioning the population $U$ in specified strata to obtain each domain as a union of the entire strata and $(ii)$ defining a uniform inclusion probability at the stratum level. For instance, consider the case in which the domains of interest are the regions and the classes of the Nace-activity codes separately. In this case, we may individuate strata by cross-classifying the modalities of the two variables region and Nace-activity code.

Let $U_{[h]}(h = 1, ..., H)$ be the generic stratum of $N_{[h]}$ units, being $U = \bigcup_{h=1}^{H} U_{[h]}$ and $N = \sum_{h=1}^{H} N_{[h]}$. Since we may obtain each domain as a union of the entire strata, we may introduce the stratum *domain* membership variable $\lambda_{d[h]}$ where $\lambda_{d[h]} = 1$ if $U_{[h]} \in U_d$ and $\lambda_{d[h]} = 0$, otherwise. If $\lambda_{d[h]} = 1$, we have $\sum_{i \in U_{[h]}} \lambda_{di} = N_{[h]}$.

We assign a uniform inclusion probability to units in stratum $U_{[h]}$

(4.1) $\quad \pi_i = \dfrac{n_{[h]}}{N_{[h]}} \quad$ for $i \in N_{[h]}$,

where $n_{[h]}$ is the expected sample size of stratum $U_{[h]}$.

Under the just illustrated sampling design, we may reformulate OP 3.2 as

(4.2) $\begin{cases} Min \left( \sum_{h=1}^{H} C_{[h]} n_{[h]} \right) \\ N_d^2 \dfrac{\sigma_{(v)u}^2 \sigma_{(v)}^2}{\left( \sum_{h=1}^{H} n_{[h]} \lambda_{d[h]} \right) \sigma_{(v)u}^2 + \sigma_{(v)}^2} \leq g1_{(v)d}^* \quad \text{for } v = 1, ..., V; d = 1, ..., D \\ 0 < n_{[h]} \leq N_{[h]} \quad\quad\quad\quad\quad\quad\quad\quad \text{for } h = 1, ..., H \end{cases}$

---

[4] http://github.com/stevengj/nlopt

where $C_{[h]}$ is the uniform unit cost for surveying the generic unit in the stratum $U_{[h]}$.

Since $H \ll N$ in most situations, Problem 4.2 is computationally much more straightforward than Problem 3.2 because it considers only $H$ unknown values.

After the definition of the sample sizes $n_{[h]}$ as the solution of Problem 4.2, we select the sample with the cube algorithm, with balancing variables given by

$$(4.3) \quad \boldsymbol{b}_i = \frac{n_{[h]}}{N_{[h]}} \boldsymbol{\lambda}_i \quad \text{for } i \in U_{[h]}.$$

We may reformulate Balancing Equations 2.4 as

$$(4.4) \quad \boldsymbol{n} = (n_1, \ldots, n_d, \ldots, n_D) = \sum_{h=1}^{H} \sum_{i \in U_{[h]}} \frac{n_{[h]}}{N_{[h]}} \boldsymbol{\lambda}_i.$$

Since, according to problem 4.2, the stratum sampling sizes $n_{[h]}$ may be fractional and lower than 1, the sample sizes of the strata are not strictly controlled. Nevertheless, balancing Equations 4.4 ensure that the sampling design defines the sample sizes at the aggregated domain level. We denote the introduced sampling design as an Incomplete multi-way Stratified Sampling (ISS) design (Falorsi and Righi, 2015).

We have a standard Stratified Simple Random Sampling Without Replacement (SSRSWOR) design if the sampling sizes $n_{[h]}$ are integers, with $n_{[h]} \geq 1$, and we want to control the stratum sampling sizes. In this case, we select the units with standard sampling selection techniques.

**5 Two-stage sampling**

This section examines the stratified two-stage sampling usually adopted in large-scale surveys on human populations. Consider the population partitioned in strata introduced in section 4 and the random mean Model 2.5. Suppose that the population of each stratum $U_{[h]}$ (for $h = 1, \ldots, H$) can be partitioned into $M_{[h]}$ subpopulations, called clusters. The set of clusters in the stratum is symbolically represented as $U_{I[h]} = \{U_{[h]1}, \ldots, U_{[h]i}, \ldots, U_{[h]M_{[h]}}\}$. Cluster $U_{[h]i}$ has

$N_{[h]i}$ final units, being $N = \sum_{h=1}^{H}\sum_{i=1}^{M_{[h]}} N_{[h]i}$. Suppose that the quantities $N_{[h]i}$ ($h = 1, \ldots, H$; $i = 1, \ldots; M_{[h]}$) are available in the sampling frame.

We select the sample $S$ using a two-stage sampling design without replacement, where we carry out the first stage of sampling with an ISS design. A first-stage sample, $S_{I[h]}$ (for $h = 1, \ldots, H$), of size $m_{[h]}$ is selected without replacement from $U_{I[h]}$, with inclusion Probabilities $\pi_{I[h]i}$ ($i = 1, 2 \ldots, M_{[h]}$) Proportional to Size (PPS):

(5.1) $\pi_{I[h]i} = m_{[h]} \dfrac{N_{[h]i}}{N_{[h]}}$.

A second-stage sample, $S_{II[h]i}$, of fixed size, $\bar{n}$, is selected from sample cluster $U_{[h]i}$ by drawing the units without replacement with equal probabilities. The second-stage inclusion probability, $\pi_{[h]IIi}$, of people in the sampled cluster $U_{[h]i}$ is

(5.2) $\pi_{II[h]i} = \dfrac{\bar{n}}{N_{[h]i}}$.

The final inclusion probability of the final unit $j$ being selected from cluster $U_{[h]i}$ is

(5.3) $\pi_{[h]ij} = \pi_{I[h]i}\pi_{II[h]i} = m_{[h]} \dfrac{N_{[h]i}}{N_{[h]}} \dfrac{\bar{n}}{N_{[h]i}} = m_{[h]} \dfrac{\bar{n}}{N_{[h]}}$

The sampling is *self-weighing*, (Kish, 1966) at the stratum level in the sense that all the units in $U_{[h]}$ have an equal probability of being selected, irrespective of their cluster. The *self-weighing* property defines a sampling design that avoids the negative impact of the variability of the sampling weights.

The first stage sampling sizes $m_{[h]}$ are defined as solution of the following OP

(5.4) $\begin{cases} Min\left(\sum_{h=1}^{H} C_{[h]} m_{[h]} \bar{n}\right) \\ N_d^2 \dfrac{\sigma_{(v)u}^2 \sigma_{(v)}^2}{\left(\sum_{h=1}^{H} m_{[h]} \lambda_{d[h]} \bar{n}\right)\sigma_{(v)u}^2 + \sigma_{(v)}^2} \le g1_{(v)d}^* & \text{for } v = 1, \ldots, V; d = 1, \ldots, D, \\ 0 < m_{[h]} \le M_{[h]} & \text{for } h = 1, \ldots, H \end{cases}$

In large-scale surveys, $\bar{n}$ is often defined based on cost and operational conditions. For example, it may not be economically feasible to involve a primary unit in the sampling, collecting data on less than 10 final units. More complex solutions, not examined here, could define an optimal value of $\bar{n}$ based on the ratio of sampling cost per first-stage unit over the cost of a final unit (Cochran 1977).

We select the first stage sample with the cube algorithm, with balancing variables given by

(5.5) $\quad \boldsymbol{b}_i = m_{[h]} \dfrac{N_{[h]i}}{N_{[h]}} \boldsymbol{\lambda}_i \quad \text{for } i \in U_{[h]}.$

Let $\boldsymbol{m} = (m_1, \ldots, m_d, \ldots, m_D)$ be the vector of the domain first stage sampling sizes such that

(5.6) $\quad \boldsymbol{n} = \bar{n}\,\boldsymbol{m}$

We define the balancing equations of the first stage sampling as

(5.7) $\quad \boldsymbol{m} = \displaystyle\sum_{h=1}^{H} \sum_{i \in U_{[h]}} m_{[h]} \dfrac{N_{[h]i}}{N_{[h]}} \boldsymbol{\lambda}_i.$

The first stage adopts an ISS design, since, the first-stage stratum sampling sizes $m_{[h]}$ may be fractional and lower than 1. The first sample sizes of the strata are not strictly controlled. But balancing Equations 5.7 ensure that the sampling design fixes the sample sizes at the aggregated domain level.

At the end of this Section, we note that by combining the sampling designs introduced in Sections 4 and 5, we cover the sampling designs of most of the social surveys on households conducted at the international level. Indeed, these surveys adopt a combined sample selection scheme: for clusters larger than a specific size, a stratified selection scheme is adopted where the clusters coincide with the strata. For the remaining clusters, a two-stage sampling design is adopted.

## 6. Uncertainty on domain membership variables

Consider the population partitioned in strata introduced in section 4 and the random mean Model 2.5. Let us suppose that the quantities $\lambda_{di}$ are not available on the sampling frame. On the other hand, we assume that the sampling researcher has sufficiently accurate information to define a model, on the probability that the unit $i$ belongs to the domain $U_d$. Let us then suppose that this model is uniform at the stratum level, being for $i \in U_{[h]}$

(6.1) $\Pr(\lambda_{di} = 1 | i \in U_{[h]}) = E_\zeta(\lambda_{di} | i \in U_{[h]}) = \phi_{di} = \phi_{d[h]},$

$V_\zeta(\lambda_{di} | i \in U_{[h]}) = \phi_{d[h]}(1 - \phi_{d[h]})$

Practically, $\phi_{d[h]}$ may be the proportion of units in the stratum belonging to the domain derived from previous surveys.

Adopting the sampling design illustrated in Section 4, the expected value of $g1_{(v)d}$, under model 6.1, maybe linearly approximated as

(6.2) $E_\zeta(g1_{(v)d}) \cong \left(\sum_{h=1}^{H} N_{[h]}\phi_{d[h]}\right)^2 \frac{\sigma_{(v)u}^2 \sigma_{(v)}^2}{\left(\sum_{h=1}^{H} n_{[h]}\phi_{d[h]}\right)\sigma_{(v)u}^2 + \sigma_{(v)}^2}.$

A simple solution for sampling in this informative context is to establish the sample sizes $n_{[h]}$ so that the expected values $n_{[h]}\phi_{d[h]}$, ensure a sufficient sample size to guarantee the approximated expected values of $g1_{(v)d}$ under 6.1 are lower than the thresholds $g1_{(v)d}^*$.

Considering the $\phi_{d[h]}$ is known, OP 4.2 can be formulated as

(6.3) $\begin{cases} Min\left(\sum_{h=1}^{H} C_{[h]} n_{[h]}\right) \\ \left(\sum_{h=1}^{H} N_{[h]}\phi_{d[h]}\right)^2 \frac{\sigma_{(v)u}^2 \sigma_{(v)}^2}{\left(\sum_{h=1}^{H} n_{[h]}\phi_{d[h]}\right)\sigma_{(v)u}^2 + \sigma_{(v)}^2} \leq g1_{(v)d}^* \quad \text{for } v = 1, \dots, V; d = 1, \dots, D \\ 0 < n_{[h]} \leq N_{[h]} \quad \text{for } h = 1, \dots, H \end{cases}$

The sample is selected with balancing Equations

(6.4) $n = \sum_{h=1}^{H} \sum_{i \in U_{[h]}} \frac{n_{[h]}}{N_{[h]}} \phi_i$

being $\boldsymbol{\phi}_i = (\phi_{1i}, \dots, \phi_{di}, \dots, \phi_{Di})'$.

## 7. General small area unit-level models

The model may be formally expressed as

(7.1) $y_{(v)di} = \boldsymbol{x}'_{di}\boldsymbol{\beta}_{(v)} + \boldsymbol{z}'_{di}\boldsymbol{u}_{(v)} + e_{(v)di}.$

where $y_{(v)di} = y_{(v)i}\lambda_{di}$ and $e_{(v)di} = e_{(v)i}\lambda_{di}$ denote the values of target variable and of random error; $\boldsymbol{x}_{di} = \boldsymbol{x}_i\lambda_{di}$ and $\boldsymbol{z}_{di} = \boldsymbol{z}_i\lambda_{di}$ are two covariate vectors available for the $i$-th unit while $\boldsymbol{\beta}_{(v)}$ and $\boldsymbol{u}_{(v)}$ are the correspondent *fixed* and *random* effects vectors respectively of dimension $\Delta$ and $D$. $\boldsymbol{x}_{di} = [x_{di,g}]$ may include quantitative and/or qualitative variables, where $x_{di,g}$ ($g = 1, \dots, \Delta$) denotes the generic element of $\boldsymbol{x}_{di}$. The generic element $z_{di,l}$ ($l = 1, \cdots, D$) of $\boldsymbol{z}_{di}$, is equal to 1 when $l = d$ while $z_{di,l} = 0$ otherwise. The vector of random errors $\boldsymbol{e}_{(v)} = [e_{(v)di}]$, is supposed to have an i.i.d. multi-Normal, $MN_N(\boldsymbol{0}, \boldsymbol{R}_{(v)})$, distribution with vector of means equal to $\boldsymbol{0}_N$ and variance covariance matrix $\boldsymbol{R}_{(v)} = \boldsymbol{R}(\sigma^2_{(v)})$ with a diagonal structure dependent on the unknown variance component $\sigma^2_{(v)}$. In particular, $\boldsymbol{R}_{(v)} = \sigma^2_{(v)}\boldsymbol{W}$ and $\boldsymbol{W} = diag_{i=1}^N\{w_i\}$ being $w_i$ a known quantity assigned to $i$-th unit. Vector $\boldsymbol{u}_{(v)}$ is supposed to have an i.i.d. multi-Normal, $MN_D(\boldsymbol{0}, \boldsymbol{G}_{(v)})$, distribution with vector of means equal to $\boldsymbol{0}_D$ and a block diagonal structure for variance covariance matrix $\boldsymbol{G}_{(v)} = \boldsymbol{G}(\boldsymbol{\omega}_{(v)})$ dependent on the vector $\boldsymbol{\omega}_{(v)}$ of variance components $\boldsymbol{\omega}_{(v)} = [\sigma^2_{(v)}, \sigma^2_{(v)u}, \rho_{(v)}]'$. In details, $\boldsymbol{G}_{(v)} = \sigma^2_{(v)u}\boldsymbol{\Omega}_{(v)}$, where $\boldsymbol{\Omega}_{(v)} = \varphi_{(v)}\boldsymbol{\Omega}(\rho_{(v)})$, , being $\varphi_{(v)} = \sigma^2_{(v)u}/\sigma^2_{(v)}$. The structure of $\boldsymbol{\Omega}_{(v)}$ is known unless the variance component vectors $\varphi_{(v)}$ and $\rho_{(v)}$. For $\boldsymbol{\Omega}(\rho_{(v)})$ we assume three alternative models $M_1 \div M_3$ which depend on the specific internal structure of the matrix $\boldsymbol{\Omega}_{(v)} = \{\varpi_{(v)dd'}; dd' \in Par(d)\}$. The first one is the basic *ANOVA* model, in which $\rho_{(v)} = 0$ and $\boldsymbol{\Omega}_{(v)} = \varphi_{(v)}\boldsymbol{I}_D$, being $\varpi_{dd'} = 1$ when $d = d'$ and $\varpi_{dd'} = 0$ otherwise. $M_2$, assumes an AR(1) process, in which $\varpi_{(v)dd'} =$

$[(1 - \rho_{(v)})]^{-1}\rho_{(v)}^{|d-d'|}$. $M_3$, assumes a process depending on a spatial distance function, $s(d, d')$, and $\varpi_{(v)dd'} = \{1 + \delta_{d,d'} + exp^{[s(d,d')\rho_{(v)}^{-1}]}\}^{-1}$, being $\delta_{d,d'} = 1$ for $d = d'$ and equal to zero otherwise. On the basis of the hypothesis of the model for $\boldsymbol{e}_{(v)}$ and $\boldsymbol{u}_{(v)}$, vector $\boldsymbol{y}_{(v)} = [y_{(v)di}]$ has a multi-Normal, $MN_N(\boldsymbol{X\beta}_{(v)}, \boldsymbol{V}_{(v)})$ distribution with vector of means $\boldsymbol{X\beta}_{(v)}$ and variance covariance matrix $\boldsymbol{V}_{(v)} = \sigma_{(v)}^2 \boldsymbol{\Sigma}_{(v)}$ being $\boldsymbol{\Sigma}_{(v)} = \boldsymbol{W} + \boldsymbol{Z}\boldsymbol{\Omega}_{(v)}\boldsymbol{Z}'$.

It is convenient to introduce the matrix formulation of model 7.1 with reference to the $N_d$ population units of $d-th$ ($d = 1, \cdots, D$) target domain, given by

(7.2) $\boldsymbol{y}_{(v)d} = \boldsymbol{X}_d\boldsymbol{\beta}_{(v)} + \boldsymbol{Z}_d\boldsymbol{u}_{(v)} + \boldsymbol{e}_{(v)}$

being: $\boldsymbol{y}_{(v)d} = [y_{(v)di}]$; $\boldsymbol{e}_{(v)d} = [e_{(v)di}]$; $\boldsymbol{X}_d = [\boldsymbol{x}'_{di}]$; $\boldsymbol{Z}_d = [\boldsymbol{z}'_{di}]$.

Under the unit – level LMM 7.1, the expression of the target population total is given by

(7.3) $Y_{(v)d} = y_{(v)d}^+ = \boldsymbol{x}_d'^+ \boldsymbol{\beta}_{(v)} + \boldsymbol{z}_d'^+ \boldsymbol{u}_{(v)}$

where $y_{(v)d}^+ = \sum_{i \in U} y_{(v)di}$, $\boldsymbol{x}_d^+ = \sum_{i \in U} \boldsymbol{x}_{di}$ and $\boldsymbol{z}_d^+ = \sum_{i \in U} \boldsymbol{z}_{di}$. Note that $\boldsymbol{z}_d^+ = N_d \boldsymbol{\xi}_d$, where $\boldsymbol{\xi}_d = [\xi_{dk}]$ is a vector of order $D$ in which $\xi_{dk} = 1$ when $k = d$ and $\xi_{dk} = 0$ otherwise.

The vector $\boldsymbol{y}_{(v)d}$ can be partitioned as $\boldsymbol{y}_{(v)d} = [\boldsymbol{y}'_{(v)ds}, \boldsymbol{y}'_{(v)dr}]'$ after the sample is selected and observed with the subscripts of "$s$" and "$r$" corresponding to sample and non-sample population units respectively. These subscripts will be used to denote conformable partitions of vectors and matrices. Then after the sample is observed, expression 7.3 may be re-written as

(7.4) $Y_{(v)d} = y_{(v)ds}^+ + \boldsymbol{x}_{dr}'^+ \boldsymbol{\beta}_{(v)} + \boldsymbol{z}_{dr}'^+ \boldsymbol{u}_{(v)}$.

in which $\boldsymbol{z}_{dr}^+ = N_{dr}\boldsymbol{\xi}_d$ for $N_{dr} = N_d - n_d$. Assuming as known the vector variance components, $\boldsymbol{\omega}_{(v)}$, and on the basis of the vector of observed data, $\boldsymbol{y}_{(v)sd}$, we obtain the estimator $\tilde{Y}_{(v)d} = \tilde{Y}_{(v)d}(\boldsymbol{\omega}_{(v)}, \boldsymbol{y}_{(v)sd})$ known as *Best Linear Unbiased Predictor* or *BLUP* of the corresponding unknown population total $Y_{(v)d}$, given by

(7.5) $\tilde{Y}_{(v)d} = y_{(v)ds}^+ + \boldsymbol{x}_{dr}'^+ \tilde{\boldsymbol{\beta}}_{(v)} + \boldsymbol{z}_{dr}'^+ \tilde{\boldsymbol{u}}_{(v)}$.

In the following for convenience we will indicate estimator 7.5 as $\tilde{Y}_{(v)d} = \tilde{Y}_{(v)d}(\boldsymbol{\omega}_{(v)})$ where $\tilde{\boldsymbol{\beta}}_{(v)} = \tilde{\boldsymbol{\beta}}_{(v)}(\boldsymbol{\omega}_{(v)})$ is the Best Linear Unbiased Estimator, BLUE of $\boldsymbol{\beta}_{(v)}$ and $\tilde{\mathbf{u}}_{(v)} = \tilde{\mathbf{u}}_{(v)}(\boldsymbol{\omega}_{(v)})$ is the BLUP of $\mathbf{u}_{(v)}$. To give the explicit expressions of these statistics, let us denote with $\mathbf{M}_{AB} = \mathbf{A}'\mathbf{W}\mathbf{B}$ and $\mathbf{m}_{Ab} = \mathbf{A}'\mathbf{W}_s\mathbf{y}_{(v)s}$ for $\mathbf{A}$ or $\mathbf{B} = (\mathbf{X}_s, \mathbf{Z}_s)$ in which $\mathbf{X}_s = [\mathbf{X}_{sd}]$ and $\mathbf{Z}_s = [\mathbf{Z}_{sd}]$. Let us consider for example $\mathbf{M}_{ZZ}$ and $\mathbf{M}_{ZX}$ where $\mathbf{Z}_s$ has the following explicit expression $\mathbf{Z}_s = diag_{d=1}^{D}\{\mathbf{1}'_{n_d}\}$, then:

(7.6) $\mathbf{M}_{ZZ} = \mathbf{Z}_s'\mathbf{W}_s\mathbf{Z}_s = diag_{d=1}^{D}\{n_d\}$,

(7.7) $\mathbf{M}_{ZX} = \mathbf{Z}_s'\mathbf{W}_s\mathbf{X}_s = [\mathbf{x}'^+_{ds}]$

in which $\mathbf{x}'^+_{ds} = \sum_{i=1}^{n_d} \mathbf{x}'_{di}$.

Then we have:

(7.8) $\tilde{\boldsymbol{\beta}}_{(v)} = M_{XX}^{-1}(\boldsymbol{\omega}_{(v)})\mathbf{m}_{Xy}(\boldsymbol{\omega}_{(v)})$,

(7.9) $\tilde{\mathbf{u}}_{(v)} = T^*[\mathbf{m}_{Zy} - \mathbf{M}_{ZX}\tilde{\boldsymbol{\beta}}_{(v)}]$.

in which:

(7.10) $M_{XX}^{-1}(\boldsymbol{\omega}_{(v)}) = [\mathbf{M}_{XX} - \mathbf{M}'_{ZX}T^*_{(v)}\mathbf{M}_{ZX}]^{-1}$,

(7.11) $\mathbf{m}_{Xy}(\boldsymbol{\omega}_{(v)}) = [\mathbf{m}_{Xy} - \mathbf{M}'_{ZX}T^*_{(v)}\mathbf{m}_{Zy}]$

being $T^*_{(v)} = T^*_{(v)}(\boldsymbol{\omega}_{(v)}) = \{t^*_{dd'}; d, d' = 1, ..., D\}$ a symmetric square matrix of order $D \times D$ given by:

(7.12) $T^*_{(v)} = [\mathbf{M}_{ZZ} + \boldsymbol{\Omega}_{(v)}^{-1}]^{-1} = [diag_{d=1}^{D}\{n_d\} + \boldsymbol{\Omega}_{(v)}^{-1}]^{-1}$,

in which $\boldsymbol{\Omega}_{(v)} = \boldsymbol{\Omega}_{(v)}(\boldsymbol{\omega}_{(v)})$. The *BLUP* estimator 7.5 depends on the vector of variance components $\boldsymbol{\omega}_{(v)}$. Therefore, by replacing in 7.8 and 7.9 $\boldsymbol{\omega}_{(v)}$ with an estimator $\hat{\boldsymbol{\omega}}_{(v)} = \hat{\boldsymbol{\omega}}_{(v)}(\mathbf{y}_{(v)s})$, for $\mathbf{y}_{(v)s} = [\mathbf{y}_{(v)sd}]$, we obtain the *two-stage* (or *plug-in*) estimator known as *Empirical Best Linear Unbiased Predictor* or *EBLUP* given by $\hat{Y}_{(v)d} = \hat{Y}_{(v)d}(\hat{\boldsymbol{\omega}}_{(v)})$. The estimates of the vector $\boldsymbol{\omega}_{(v)}$ can be obtained both by means of the *ML* or through *REML* estimators. The estimation of the variance components is obtained using an iterative procedure because the estimating equations for the variance components do not have an analytical solution and must therefore be solved through numerical procedures. The MSE of *BLUP* estimator 7.5 is

(7.13) $MSE[\tilde{Y}_{(v)d}] = g1_{(v)d}(\boldsymbol{\omega}_{(v)}) + g2_{(v)d}(\boldsymbol{\omega}_{(v)})$,

where $g1_{(v)d} = g1_{(v)d}(\boldsymbol{\omega}_{(v)})$ and $g2_{(v)d} = g2_{(v)d}(\boldsymbol{\omega}_{(v)})$ are given by

(7.14) $g1_{(v)d} = \frac{\sigma_{(v)}^2}{n_d} \mathbf{z}'^+_{dr}\boldsymbol{\Gamma}_{(v)}\mathbf{z}^+_{dr} = N_{dr}^2 \frac{\sigma_{(v)}^2}{n_d}\gamma_d \cong N_d^2 \frac{\sigma_{(v)}^2}{n_d}\gamma_{(v)d}$,

(7.15) $g2_{(v)d} = \sigma_{(v)}^2 [\mathbf{x}'^+_{dr} - n_d^{-1}\mathbf{z}'^+_{dr}\boldsymbol{\Gamma}_{(v)}\mathbf{M}_{ZX}]\mathbf{M}_{XX}^{-1}(\boldsymbol{\omega}_{(v)})[\mathbf{x}'^+_{dr} - n_d^{-1}\mathbf{z}'^+_{dr}\boldsymbol{\Gamma}_{(v)}\mathbf{M}_{ZX}]'$

where $\gamma_{(v)d}$ is the $d - th$ element on the main diagonal of $\boldsymbol{\Gamma}_{(v)}$ in which:

(7.16) $\boldsymbol{\Gamma}_{(v)} = [diag_{d=1}^D\{n_d\}] [diag_{d=1}^D\{n_d\} + \boldsymbol{\Omega}_{(v)}^{-1}]^{-1} = \{\gamma_{dd'}; d, d' = 1,..,D\}$,

being $\gamma_{dd'} = n_d t_{dd'}$  $\bar{\mathbf{x}}'^+_{ds} = \frac{1}{n_d}\sum_{i=1}^{n_d} \mathbf{x}'_{di}$.

Considering 7.7 and $\mathbf{z}^+_{dr} = N_{dr}\boldsymbol{\xi}_d$, $g_2$ may be simplified as

$g2_{(v)d} = \sigma_{(v)}^2 \left[\mathbf{x}'^+_{dr} - N_{dr}\sum_{d'}\bar{\mathbf{x}}'^+_{d's}\gamma_{dd'}\right]\mathbf{M}_{XX}^{-1}(\boldsymbol{\omega}_{(v)})\left[\mathbf{x}'^+_{dr} - N_{dr}\sum_{d'}\bar{\mathbf{x}}'^+_{d's}\gamma_{dd'}\right]'$.

Under the ANOVA model in which $\boldsymbol{\Omega}_{(v)} = \varphi_{(v)}\mathbf{I}_D$ and $\gamma_{dd'} = 0$ for $d \neq d'$, we obtain expression 2.10 of the simple *random-mean* model where we have $\mathbf{x}'_{di} \equiv 1$ and $w_i = 1$ $\forall i$. The term $g2_{(v)d}$ is of a minor order and it vanishes for large sample sizes $n$. Therefore, we can define the vector of inclusion probabilities $\boldsymbol{\pi}$ focusing on the term $g1_{(v)d}$ .

### 7.1 *The fixed-point algorithm*

If we consider Expression 7.14, we see that the OP cannot be defined as given in 3.2 since the $\pi_i$ values are implicitly defined as embedded in the inverse of matrix $\boldsymbol{\Gamma}_{(v)}$. In contrast, expression 3.2 defines the inclusion probabilities explicitly in the constraints' expressions. Therefore, for this case we have to individuate a different algorithmic solution.

Considering 7.16, we see that the term $\gamma_{(v)d}$ of 7.14, which is the $d-th$ element on the main diagonal of $\mathbf{\Gamma}_{(v)}$, may be expressed as function of the inclusion probabilities

(7.17) $\quad \mathbf{\Gamma}_{(v)} = \left[ diag_{d=1}^{D} \left\{ \sum_{i \in U} \pi_i \lambda_{di} \right\} \right] [diag_{d=1}^{D} \{ \sum_{i \in U} \pi_i \lambda_{di} \} + \mathbf{\Omega}_{(v)}^{-1}]^{-1}.$

Expression 7.17 represents the basis for setting up a fixed-point algorithm to define the optimal vector $\boldsymbol{\pi}$ of inclusion probabilities described below.

Let $t = 0,1, ....$ denote the generic iteration and let $\boldsymbol{\pi}^t = (\pi_1^t, ..., \pi_i^t, ..., \pi_N^t)'$ the value of the vector $\boldsymbol{\pi}$ at iteration $t$. Let $\gamma_{(v)d}(\boldsymbol{\pi}^t)$ be the value of $\gamma_{(v)d}$ as a function of the vector $\boldsymbol{\pi}^t$, where $\gamma_{(v)d}(\boldsymbol{\pi}^t)$ is the element in the main diagonal of

(7.18) $\quad \mathbf{\Gamma}_{(v)} = \left[ diag_{d=1}^{D} \left\{ \sum_{i \in U} \pi_i^t \lambda_{di} \right\} \right] [diag_{d=1}^{D} \{ \sum_{i \in U} \pi_i^t \lambda_{di} \} + \mathbf{\Omega}_{(v)}^{-1}]^{-1}.$

*Step 0. Initialization*

We determine the vector $\boldsymbol{\pi}^0$ as $\boldsymbol{\pi}^0 = \{\pi; 1, ..., N\}$ where $0 < \pi < 1$ is a fixed value. We compute $\gamma_{(v)d}(\boldsymbol{\pi}^0)$ according to (7.18)

*Step 1. Calculus.* At iteration $t = 1,2, ....$ we solve the following OP

(7.19) $\quad \begin{cases} Min \left( \sum_{i \in U} C_i \pi_i \right) \\ N_d^2 \frac{\sigma_{(v)}^2}{(\sum_{i \in U} \pi_i^t \lambda_{di})} \leq g1_{(v)d}^* \frac{1}{\gamma_{(v)d}(\boldsymbol{\pi}^{t-1})} (v = 1, ..., V; d = 1, ..., D)' \\ 0 < \pi_i^t \leq 1 \end{cases}$

taking the values $\gamma_{(v)d}(\boldsymbol{\pi}^{t-1})$ as fixed.

*Step 2. Fixed-point iteration.* Assigning $\varepsilon$ a small value, (e.g. $\varepsilon = 5$), if

(7.20) $\quad \sum_{i=1}^{N} |\pi_i^t - \pi_i^{t-1}| \leq \varepsilon,$

the iterations stop. Otherwise, we update the values $\gamma_{(v)d}(\boldsymbol{\pi}^t)$ and iterate step 2 for $t = t + 1$, until convergence.

The Fixed-point algorithm is quite general and can also be applied to OP 3.2, but a limitation could be the execution time. The Appendix demonstrates the algorithm converges to a fixed-point solution for Model 3.2. We empirically proved the convergence in Section 8. We even see in Section 8 that the fixed-point algorithm is slightly more efficient than that presented in Section 3. From the computational point of view, the algorithm is much slower than the version presented previously. The following section will also compare the two algorithms on the time performance. Practically, when the computation times of the fixed point become too long, the other version could be used because if it converges, it guarantees a good approximation of the results obtained through the fixed-point version.

## 8 Experimental results

### 8.1 *Data description*

We tested the proposed approach using the data frames "*data_s*" and "*univ*", coming from the R package MIND (D'Alò et al., 2021), MIND is an R Package implementing multivariate models for small areas at unit level such as those described in section 7. The *data_s* is a sample of 10,000 individuals and 104 municipalities selected from *univ*, which includes a population of 514,320 individuals and 333 municipalities. We considered a subset of the sample data referred to 49 municipalities to reduce computational time. The sample record for each individual has the following variables:

- Sex by age class (*Sexage*, 14 categories).
- Educational level (*Edu*, six levels),
- Nationality (*Fore*, 1=italian, 2=otherwise),
- Municipality code (M*un*, 49 codes),
- Employment status (*Emp*, 1=employed and 0=otherwise).

We randomly separate the 49 municipalities in two macro-strata: we assigned 24 municipalities to one of two, say macro-strata 1, while we set the remaining 25 municipalities to the other macro-stratum, say macro-stratum 2.

After this process, the dataset has two types of overlapping domains. The first type, denoted as partition A, has 49 domains, each corresponding to a municipality. The second partition, marked as partition B, has two domains, each of which is a macro-stratum. The domains of the B partition are an aggregation of type A partition domains.

### 8.2 *The parameter setting*

The optimization problem (4.2) needs some input parameters. The variance components $\sigma^2_{(v)u}$ and $\sigma^2_{(v)}$ have been estimated by means of the model

$$y_{(v)i}\lambda_{di} = \alpha + \boldsymbol{\beta}'_1 \mathbf{x}_{1i} + \boldsymbol{\beta}'_2 \mathbf{x}_{2i} + \boldsymbol{\beta}'_3 \mathbf{x}_{3i} + u_{(v)d} + e_{(v)i},$$

where $\lambda_{di}$ (with $d = 1, \ldots, 51$) is the membership domain variable; $\mathbf{x}'_{1i} = (x_{1,1i}, \ldots, x_{1,28i})'$ is the vector of the dummy variables of the *Sexage* variable, being $x_{1,ji}$ (with $j = 1, \ldots, 14$) equal to 1 when unit $i$ has the $j$th class of *Sexage*, and zero otherwise; $\mathbf{x}'_{2i} = (x_{2,1i}, \ldots, x_{2,6i})'$ is the vector of the dummy variables of the *Edu* variable, being $x_{2,ji}$ (with $j = 1, \ldots, 6$) equal to 1 when unit $i$ has the $j$th educational level of *Edu*, and zero otherwise; $\mathbf{x}'_{3i} = (x_{3,1i}, x_{3,2i})'$ is the vector of the dummy variables of the *Fore* variable, being $x_{3,ji}$ (with $j = 1,2$) equal to 1 when unit $i$ has the $j-th$ status of *Fore*, and zero otherwise; $\boldsymbol{\beta}_1$, $\boldsymbol{\beta}_2$ and $\boldsymbol{\beta}_3$ are respectively the vectors of intercepts of $\mathbf{x}_{1i}$, $\mathbf{x}_{2i}$ and $\mathbf{x}_{3i}$ being $\boldsymbol{\beta}'_k = (\beta_{k1}, \ldots, \beta_{kJ})'$ (with $k = 1, \ldots, 3$) and $J$ equal to the number of dummy variables of the $k$th variable; $u_{(v)d}$ is the municipality random intercept; $e_{(v)i}$ is the random error. We estimate the model on the data and obtain the two values $\sigma^2_{(v)u} = 0.0005$ and $\sigma^2_{(v)} = 0.1958$.

The optimization problem needs the precision threshold given $g1^*_{(v)d}$. The $g1^*_{(v)d}$ is a real number spanning a vast set of values. Therefore it can be complex to determine an understandable value. Therefore, we define a relative standard error given by the expression

$$R^*_{(v)d} = \sqrt{g1^*_{(v)d}/Y_{(v)d}},$$

that is, in general (for positive target parameter), a real number in an interval of values greater than zero and smaller than 1. The use of $R^*$ entails a hint of $Y_{(v)d}$ (that is the target parameter).

Then, setting $R^*_{(v)d} = t_{(v)d}$ we have $g1^*_{(v)d} = (t_{(v)d} Y_{(v)d})^2$.

In our application. we use a value of $Y_{(v)d} \cong 0.28 \, N_d$.

Finally, we set initial domain sample sizes $\mathbf{n}'_0 = (n_{01}, \ldots, n_{0d}, \ldots, n_{0D})'$ being $D = 49 + 2$. One of the below experiments shows that $\mathbf{n}_0$ does not affect the solution.

### 8.2.1 The experimental settings

We study the proposed sample allocation method varying the precision thresholds and the initial sample sizes. We compare the allocation given by the (4.2) with the proportional allocation to the population size in each domain. The proportional allocation is the expected domain allocation when the domain sample sizes are uncontrolled. The overall sample size of proportional sampling allocation is set equal to the overall sample size of the proposed sample allocation.

Finally, we denote by $R_{(v)d}$ the obtained relative error, being $R_{(v)d} = \sqrt{g1_{(v)d}/Y_{(v)d}}$, where $g1_{(v)d}$ is obtained with a given $n_d$ value. In particular, we denote by $R^{opt}_{(v)d}$ the obtained relative error with the $n^{opt}_d$ computed with the proposed sample allocation. $R^{pro}_{(v)d}$ is the obtained relative error with the $n^{pro}_d$ computed with the proportional sample allocation.

### 8.3 Experiments

### 8.3.1 Experiment 1

The first experiment sets two different precision thresholds, $R^*$, for the A and B types of domain. For each of the 49 domains of partition A we set $R^*_{(v)d} = 0.07$. For the 2 domains of partition B, we set $R^*_{(v)d} = 0.05$. Eventually $\mathbf{n}'_0 = \mathbf{1}'$ where $\mathbf{1}$ is a vector of 1s of size 51.

**Tab. 8.1** – First experiment: the sample allocations in the domains of partition A

| Domains classes by population size | | | Proposed Allocation | | | | | Proportional Allocation | | Eff= $R^{pro}_{(v)d}/R^{opt}_{(v)d}$ |
|---|---|---|---|---|---|---|---|---|---|---|
| Domain quartile^ | Average $N_d$ | Average $g1^*_{(v)d}$ | Average of $n_d$ | Average of the realized $R^{opt}_{(v)d}$ | $RAP^{opt}_{(v)d} = R^{opt}_{(v)d}/R^*_{(v)d}$ | | | Average of $n_d$ | Average of the realized $R^{pro}_{(v)d}$ | |
| | | | | | Q1 | Average | Q3 | | | |
| Min-Q1 | 801.17 | 290.72 | 115.67 | 0.070 | 1.00 | 1.00 | 1.00 | 28.17 | 0.077 | 1.10 |
| Q1-Q2 | 1,520.08 | 938.45 | 145.42 | 0.070 | 1.00 | 1.00 | 1.00 | 53.50 | 0.075 | 1.10 |
| Q2-Q3 | 2,218.75 | 1957.50 | 149.92 | 0.069 | 1.00 | 1.00 | 1.00 | 78.08 | 0.074 | 1.07 |
| Q3-Max | 9,924.54 | 129,175.0 | 117.08 | 0.069 | 1.00 | 1.00 | 1.00 | 349.23 | 0.062 | 0.89 |
| All domains | 3,744.88 | 35051.34 | 131.71 | 0.069 | 1.00 | 1.00 | 1.00 | 131.78 | 0.072 | 1.04 |

^Q1: first quartile; Q2: second quartile (median); Q3: third quartile.

**Tab. 8.2** – First experiment: the sample allocations in the domains of partition B

| Domains classes by population size | | | Proposed Allocation | | | Proportional Allocation | | Eff= $R^{pro}_{(v)d}/R^{opt}_{(v)d}$ |
|---|---|---|---|---|---|---|---|---|
| Domain | $N_d$ | $g1^*_{(v)d}$ | $n_d$ | Realized $R^{opt}_{(v)d}$ | $RAP^{opt}_{(v)d} = R^{opt}_{(v)d}/R^*_{(v)d}$ | $n_d$ | Realized $R^{pro}_{(v)d}$ | |
| b1 | 47,548 | 465,041.31 | 2,610 | 0.030 | 0.600 | 1,672 | 0.036 | 1.21 |
| b2 | 13,5951 | 3,801,822.49 | 3,844 | 0,024 | 0.480 | 4,782 | 0.021 | 0.90 |

Table 8.1 shows the main results of partition A, while table 8.2 shows the results of partition B. The results of partition A are summarised by the population size of the domains classified into four classes. Domains with population size: less than the first quartile (Q1); greater or equal to the Q1 and less than the second quartile (Q2); greater or equal to the Q2 and less than the third quartile (Q3); greater or equal to the Q3. Table 8.1 and 8.2 highlight that the precision threshold for the domains of partition A drives the proposed sample allocation. In fact, to satisfy the 0.05 threshold of the two domains of partition B, we need fewer records than 2,610 (first domain) and 3,844 (second domain) since the $R^{opt}_{(v)d}$, equal to 0.030 and 0.024, are mainly inferior to 0.05. Nevertheless, the optimization problem assigns these sample sizes to satisfy the constraints on the partition A domains. So, let us focus on the analysis of partition A. We point out that the proposed allocation gives sample sizes such that the obtained $R^{opt}_{(v)d}$ is close to $R^{*}_{(v)d}$. Conversely, the proportional allocation gives many more records to the large domains (Q3-Max) and a few units to the remaining domains. Table 8.1 shows the proportional allocation is inefficient for the small domains (the precision constraints are not satisfied) while it is too efficient for the large domains, in the class Q3-Max (the average of $R^{pro}_{(v)d} > R^{*}_{(v)d}$). The conclusive remark is that the proposed method oversamples the small domain according to a well-defined methodological framework. Figure 8.1 below graphically represents what we have just been described.

**Figure 8.1** - First experiment: trend of $RAP^{opt}_{(v)d} = R^{opt}_{(v)d}/R^{*}_{(v)d}$ and $RAP^{pro}_{(v)d} = R^{pro}_{(v)d}/R^{*}_{(v)d}$ by $N_d$

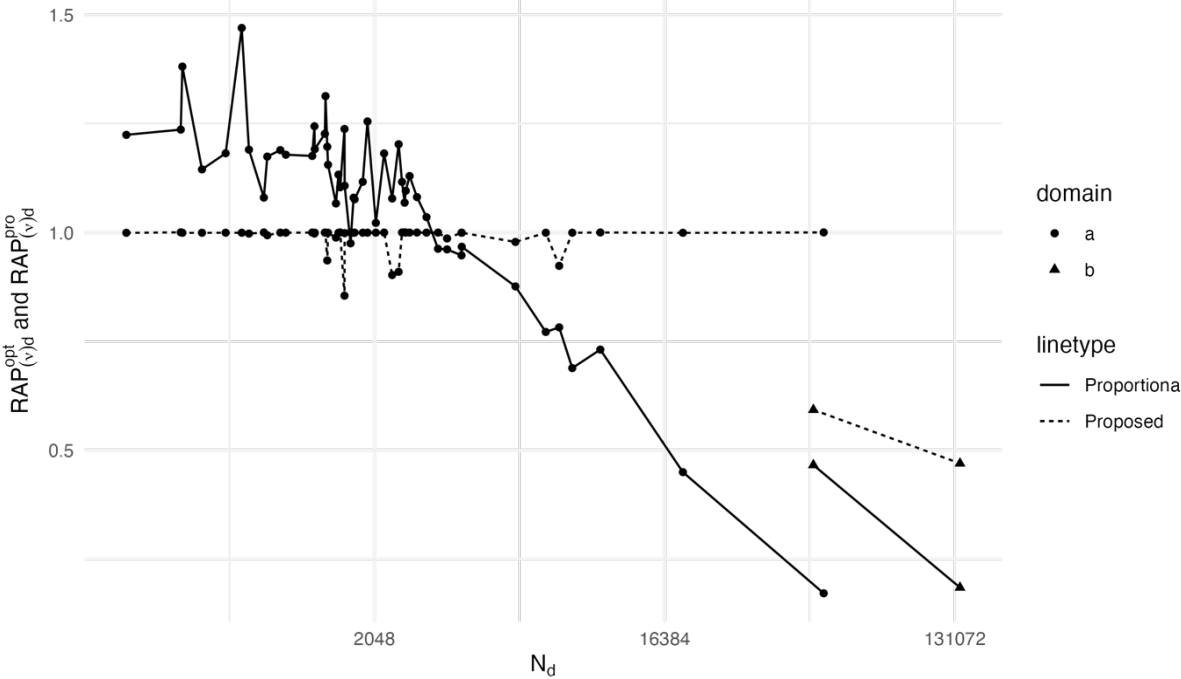

### 8.3.2 Experiment 2

The second experiment changes the precision threshold $R^*$ for the A types of domains from 0.07 to 1.00. Tables 8.3 and 8.4 show that the precision threshold for the domains of partition B drives the sample allocation. That means the proposed allocation oversamples the domains of partition A to satisfy the precision threshold on the domains of partition B. Compared to the proportional allocation, the sample sizes of the obtained relative error, $R_{(v)d}$, fully respect the precision thresholds when observing partition B. The constraints are not satisfied by the proportional allocation method. Figure 8.2 clearly show these findings.

**Tab. 8.3** - Second experiment: the sample allocations in the domains of partition A

| Domains classes by population size | | | Proposed Allocation | | | | | Proportional Allocation | | Eff= $R^{pro}_{(v)d}/R^{opt}_{(v)d}$ |
|---|---|---|---|---|---|---|---|---|---|---|
| Domain quartile^ | Average $N_d$ | Average $g1^*_{(v)d}$ | Average of $n_d$ | Average of the realized $R^{opt}_{(v)d}$ | $RAP^{opt}_{(v)d} = R^{opt}_{(v)d}/R^*_{(v)d}$ | | | Average of $n_d$ | Average of the realized $R^{pro}_{(v)d}$ | |
| | | | | | Q1 | Average | Q3 | | | |
| Min-Q1 | 801.17 | 593.30 | 52.08 | 0.075 | 0.52 | 0.59 | 0.62 | 5.00 | 0.080 | 1.06 |
| Q1-Q2 | 1,520.08 | 1,915.21 | 19.50 | 0.079 | 0.60 | 0.63 | 0.65 | 9.58 | 0.080 | 1.01 |
| Q2-Q3 | 2,218.75 | 3,994.89 | 5.92 | 0.081 | 0.63 | 0.65 | 0.69 | 14.00 | 0.080 | 0.99 |
| Q3-Max | 9,924.54 | 263,622.5 | 18.31 | 0.078 | 0.61 | 0.62 | 0.65 | 63.23 | 0.074 | 0.95 |
| All domains | 3,744.88 | 71,533.35 | 23.84 | 0.078 | 0.59 | 0.62 | 0.65 | 23.78 | 0.078 | 1.00 |

^Q1: first quartile; Q2: second quartile (median); Q3: third quartile.

**Tab. 8.4** - Second experiment: the sample allocations in the domains of partition B

| Domains classes by population size | | | Proposed Allocation | | | Proportional Allocation | | Eff= $R^{pro}_{(v)d}/R^{opt}_{(v)d}$ |
|---|---|---|---|---|---|---|---|---|
| Domain | $N_d$ | $g1^*_{(v)d}$ | $n_d$ | Realized $R^{opt}_{(v)d}$ | $RAP^{opt}_{(v)d} = R^{opt}_{(v)d}/R^*_{(v)d}$ | $n_d$ | Realized $R^{pro}_{(v)d}$ | |
| b1 | 47,548 | 465,041.31 | 583 | 0.053 | 1.05 | 303 | 0.062 | 1.19 |
| b2 | 13,5951 | 3,801,822.49 | 585 | 0.049 | 0.99 | 865 | 0.043 | 0.88 |

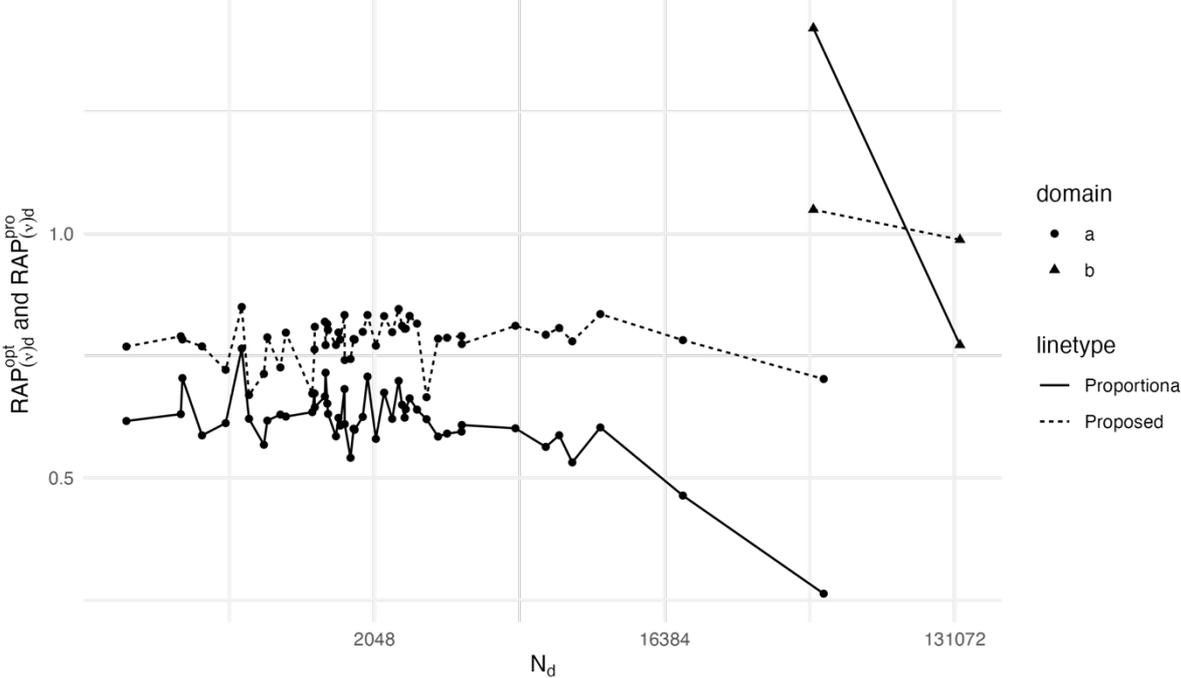

**Figure 8.2** - Second experiment: trend of $RAP_{(v)d}^{opt} = R_{(v)d}^{opt}/R_{(v)d}^{*}$ and $RAP_{(v)d}^{pro} = R_{(v)d}^{pro}/R_{(v)d}^{*}$ by $N_d$

### 8.3.3 Experiment 3

The third experiment considers the setting of experiment 1 with a different vector of initial sample size $\mathbf{n}_0$. We apply the optimization problem with the initial sample sizes ranging from 0.005 to 0.05 of the domain population. Figure 8.3 shows that the optimization process finds a solution in every case, and the speed of convergence does not seem to be affected by the starting point.

**Figure 8.3** - Rate of convergence for different initial numbers of sample units obtained as a fraction of the population size ($n_{0d}/N_d$= 0.05, 0.02, 0.01, 0.005).

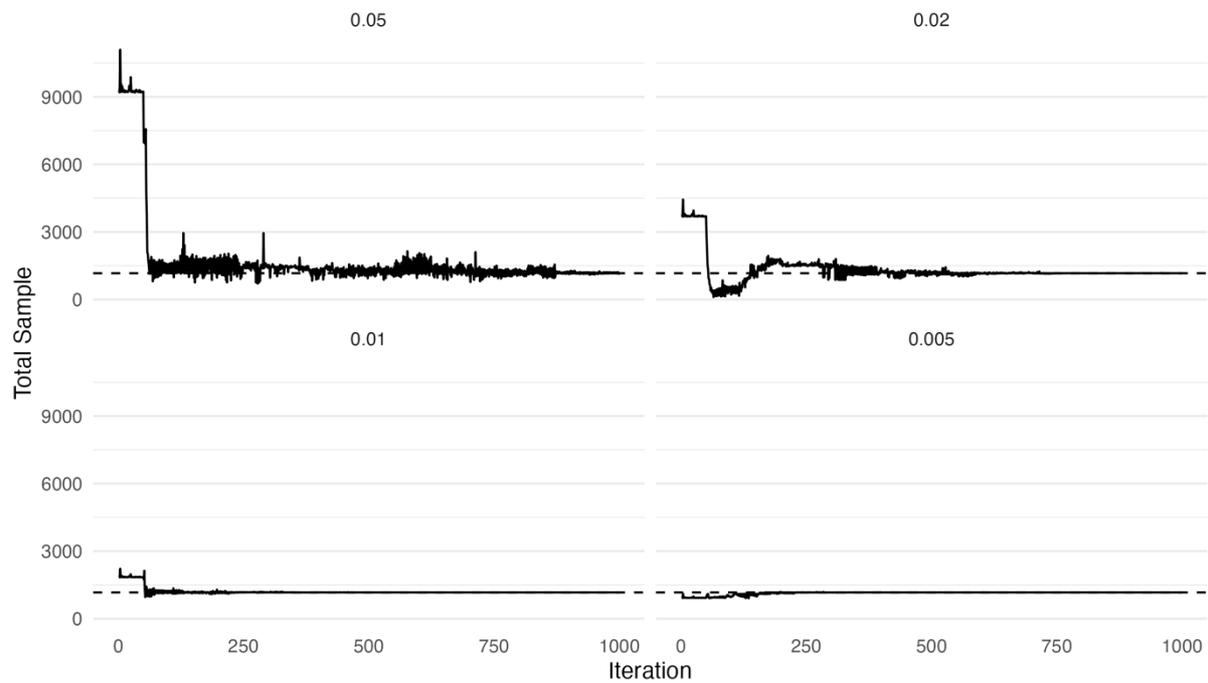

We repeat the experiment setting a uniform starting point for each domain (figure 8.4). We highlight that the different starting point brings to the convergence. For all cases, the final domain allocation vector is approximately the same.

**Figure 8.4-** Speed of convergence for different uniform numbers of sample units at the starting points ($n_{0d}$=100, 50, 10, 5).

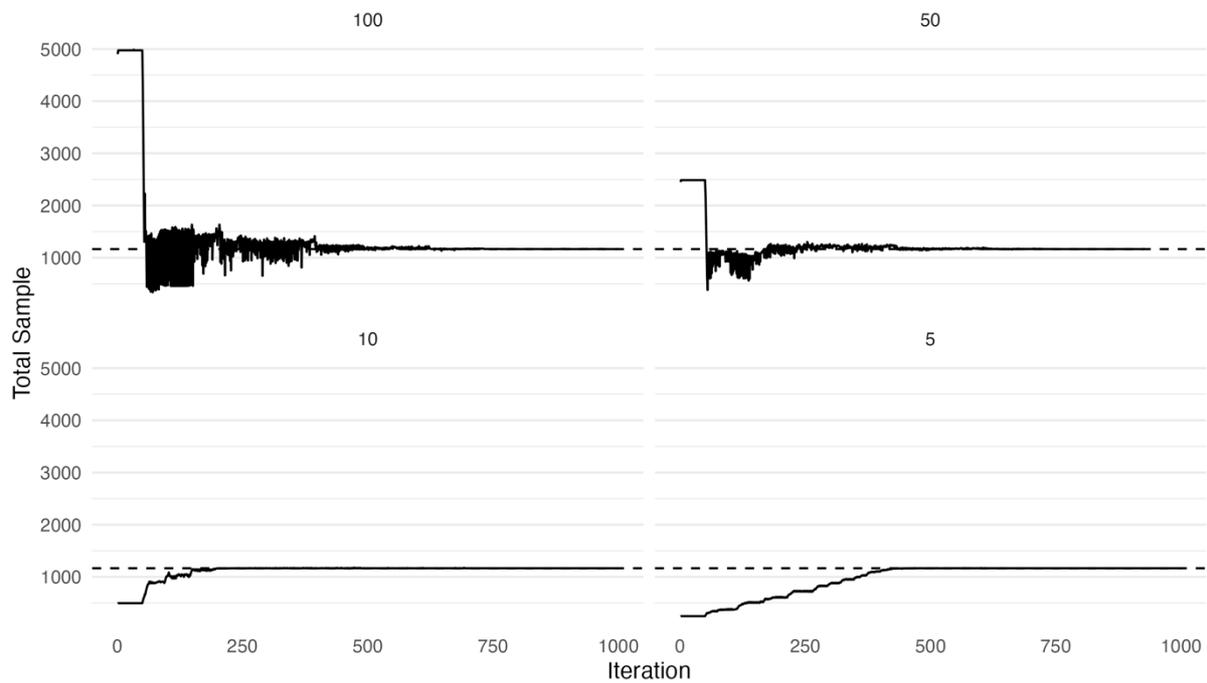

### 8.3.4 Experiment 4

The fourth experiment solves the optimization problem (4.2) by applying the fixed-point algorithm. The median computation time calculated on 15 experiments with the fixed-point algorithm is 23.20 seconds while the version used in the first experiment runs in a median time of 0.62 seconds.

The experiment uses the parameter setting of the first experiment.

The findings in table 8.5 underline a coherence with the previous experiment. The fixed-point algorithm obtains a solution very close to the precision threshold for the domains of the A partition, even though it looks more efficient than the solution of the first experiment since the overall average of $n_d$ is 118.76 while in the first experiment is 131.71.

**Tab. 8.5** - Sample allocation using the fixed-point algorithm in the domains of partition A

| Domains classes by population size | | | Proposed Allocation | | | | | Proportional Allocation | | Eff= $R^{pro}_{(v)d} / R^{opt}_{(v)d}$ |
|---|---|---|---|---|---|---|---|---|---|---|
| Domain quartile^ | Average $N_d$ | Average $g1^*_{(v)d}$ | Average of $n_d$ | Average of the realized $R^{opt}_{(v)d}$ | $RAP^{opt}_{(v)d} = R^{opt}_{(v)d}/R^*_{(v)d}$ | | | Average of $n_d$ | Average of the realized $R^{pro}_{(v)d}$ | |
| | | | | | Q1 | Average | Q3 | | | |
| Min-Q1 | 801.17 | 290.72 | 114.83 | 0.070 | 1.00 | 1.00 | 1.00 | 25.25 | 0.078 | 1.11 |
| Q1-Q2 | 1520.08 | 938.45 | 122.00 | 0.070 | 1.00 | 1.00 | 1.00 | 48.08 | 0.076 | 1.08 |
| Q2-Q3 | 2,218.75 | 1,957.50 | 131.17 | 0.070 | 1.00 | 1.00 | 1.00 | 70.42 | 0.075 | 1.07 |
| Q3-Max | 9,924.54 | 129,175,50 | 107.92 | 0.070 | 1.00 | 1.00 | 1.00 | 314.54 | 0.063 | 0.90 |
| All domains | 3,744.88 | 35,051.34 | 118.76 | 0.070 | 1.00 | 1.00 | 1.00 | 118.65 | 0.073 | 1.04 |

^Q1: first quartile; Q2: second quartile (median); Q3: third quartile.

**Tab. 8.5** - Sample allocation using the fixed-point algorithm in the domains of partition A

| Domains classes by population size | | | Proposed Allocation | | | Proportional Allocation | | Eff= $R^{pro}_{(v)d}/R^{opt}_{(v)d}$ |
|---|---|---|---|---|---|---|---|---|
| Domain | $N_d$ | $g1^*_{(v)d}$ | $n_d$ | Realized $R^{opt}_{(v)d}$ | $RAP^{opt}_{(v)d} = R^{opt}_{(v)d}/R^*_{(v)d}$ | $n_d$ | Realized $R^{pro}_{(v)d}$ | |
| b1 | 47,548 | 465,041.31 | 2,328 | 0.031 | 0.6233 | 1,507 | 0,037 | 1.20 |
| b2 | 13.5951 | 3,801,822.49 | 3,487 | 0.025 | 0.4912 | 4,310 | 0,022 | 0.91 |

## 9. Conclusions

Developing an overall strategy that deals with the Small Area Estimation (SAE) problem is a pending issue in the finite population inference. Usually, the SAE strategy leverages the specific estimation techniques moving the sampling design to the background. This paper recognizes that an accurately planned sampling design allows us to improve the precision of the SAE.

We assume that domain indicator variables are accessible for each unit in the population or if we have sufficiently accurate information regarding the probability of belonging to a given domain.

We investigate how to plan a sampling design (considering stratification and stratum sample allocation) that guarantees the sampling errors of an indirect model-based SAE of the domains will be lower than pre-specified thresholds. The sampling design minimizes the costs (expressed as the overall sample size) since the costs distinguish the SAE problems with small sample sizes and unreliable direct estimates.

We dealt with three main sampling steps. First, we defined the optimization problem for the vector of inclusion probabilities to have the expected minimum cost sampling design, ensuring the model-based estimates respect the desired model-based accuracy. This optimization problem aims to allocate the sample units in the domains of estimation efficiently.

Then, we studied the algorithmic solutions necessary to solve the optimization problem. We proposed two solutions. The first is based on a fast algorithm. It is usable when the model-based estimator relies on the simplest models. The second solution exploits a fixed-point algorithm. It is always applicable and produces a slightly efficient solution compared to the first algorithm's solution.

Finally, we proposed the sampling selection algorithms to ensure the defined sampling sizes at the small-area level. It considers a specific application of the algorithm performing a balanced random sample.

We investigated the issues mentioned above for various small area models and sampling settings, including stratification, two-stage sampling, and domain membership variable uncertainty.

Experiments on simulated data sets revealed that controlling sampling sizes is crucial for small domains to protect against inaccurate model-based estimations and ensure that the accuracy level never exceeds the established accuracy. In the experiments, the strategies neglecting the sampling design have highly heterogeneous sampling errors among the domains since some domains receive few sample units and others many more units.

The experiments show that the fast algorithm gives an approximately less efficient solution than the fixed-point algorithm. That means the fixed-point algorithm can allocate the sample units in the domains of estimates guaranteeing smaller sampling errors on average, compared to the fast algorithm. Nevertheless, according to the experiments, obtaining results quicker may cost us some inefficiency. Finally, we can apply the two algorithms sequentially: use the fast algorithm in the explorative phase to tune the input parameters, then use the fixed-point algorithm for the final solution.

**References**

Bethel, J. (1989). Sample allocation in multivariate surveys. *Survey Methodology*, 15, 47-57.

Burgard. J. P, Munnich R., and Zimmermann. T. (2014). The Impact of Sampling Designs on Small Area Estimates for Business Data. Journal of Official Statistics, Vol. 30, No. 4, 2014, pp. 749–771, http://dx.doi.org/10.2478/JOS-2014-0046

Chambers R. L., and Clark. R. G. (2012), *An introduction to Model-Based Survey Sampling with Applications*. Oxford Statistical Science Series. 37.

Cochran, W.G. (1977). *Sampling Techniques*. New York: John Wiley & Sons, Inc.

Chromy, J. (1987). Design optimisation with multiple objectives. *Proceedings of the Survey Research Methods Section, American Statistical Association*, 194-199.

D'alò, M., Falorsi, S. and Fasulo, A. (2021). MIND, an R package for multivariate small area estimation with multiple random effects., *Conference: SAE 2021*. Volume: ISBN 9788899594138.

Deville, J.-C. and Tillé, Y. (2004). Efficient Balanced Sampling: the Cube Method, *Biometrika*, 91, 893-912.

Deville, J.-C. and Tillé, Y. (2005). Variance approximation under balanced sampling, *Journal of Statistical Planning and Inference*, 128, 569-591.

Dorfman. A., Vaillant R., (2005). 26. Model-based Prediction of Finite Population Total. *Handbook of Statistics, Sample Surveys: Inference and Analysis*, Vol. 29B. ISSN: 0169-7161 2009 Elsevier B.V. DOI: 10.1016/S0169-7161(09)00223-5.

Eurostat. (2008). *NACE2. Statistical classification of economic activities in the European Community*. https://ec.europa.eu/eurostat/documents/3859598/5902521/KS-RA-07-015-EN.PDF.

Falorsi, P.D., and Righi, P. (2008). A balanced sampling approach for multi-way stratification designs for small area estimation. *Survey Methodology*, 34, 2, 223-234.

Falorsi, P. D. and Righi, P. (2015), Generalized Framework for Defining the Optimal Inclusion Probabilities of One-Stage Sampling Designs for Multivariate and Multi-domain Surveys, *Survey Methodology*, 41, 215-236.

Friedrich, U., Münnich R., and Rupp M. (2018). Multivariate Optimal Allocation with Box-Constraints. *Austrian Journal of Statistics*, 47, 33–52, doi:10.17713/ajsn.v47i2.764.

Gabler, S., M. Ganninger, and Münnich R. (2012). Optimal Allocation of the Sample Size to Strata Under Box Constraints. *Metrika*, 75, 15–161, DOI: http://dx.doi.org/ 10.1007/s00184-010-0319-3.

Grafström A, Lundström NL, Schelin L. Spatially balanced sampling through the pivotal method (2012). *Biometrics*, 68, 2, 514-20. doi: 10.1111/j.1541-0420.2011.01699.x

Grafström, A., & Lundström, N. L. (2013). Why well spread probability samples are balanced. *Open Journal of Statistics*, *3*(1), 36-41.

Grafström, A. and Tillé, Y. (2013). Doubly balanced spatial sampling with spreading and restitution of auxiliary totals". *Environmetrics,* 24, 120–131.

Keto M., Hakanen J., and Pahkinen E. (2018). Register data in sample allocations for small-area estimation, *Mathematical Population Studies*, 25:4, 184-214, DOI: 10.1080/08898480.2018.1437318.

Lehtonen, R., C.-E. Särndal, and Veijanen A. (2003). The Effect of Model Choice in Estimation for Domains, Including Small Domains, *Survey Methodology*, 29, 33–44.

Lehtonen, R., and Veijanen. A. (2009). Design-Based Methods of Estimation for Domains and Small Areas. In *Handbook of Statistics*, Vol. 29B, 219–249. New York: Elsevier.

Marker, D.A. (2001). Producing small area estimates from national surveys: Methods for minimizing use of indirect estimators. *Survey Methodology*, 27, 183-188.

Pfeffermann D. (2013). New Important Developments in Small Area Estimation. *Statistical Science*, 28, 1, 40–68, DOI: 10.1214/12-STS395

Pfeffermann. D., and Ben-hur D. (2019) Estimation of Randomization Mean Square Error in Small Area Estimation. *International Statistical Review*, May 2019. https://onlinelibrary.wiley.com/doi/10.1111/insr.12289

Powell. M.J.D. (1994) *Advances in optimization and numerical analysis*, 1994 - Springer

Rao. J.N.K. (2003) *Small Area Estimation*, Wiley New York.

Särndal, C.-E., Swensson, B., and Wretman, J., (1992) *Model Assisted Survey Sampling*. Springer-Verlag, New York, 1992. ISBN 0-387-97528-4.

Singh, M.P., Gambino, J. and Mantel, H.J. (1994). Issues and strategies for small area data. *Survey Methodology*, 20, 3-22.

Tillé, Y. and Favre, A.-C. (2005). Optimal Allocation in Balanced Sampling. *Statistics and Probability Letters*, 74, 31–37.

Tillé, Y. (2020). Sampling and estimation from finite populations. *John Wiley & Sons*.

Vaillant R., Dorfman, A.H. and Royall R.M (2000) *Finite population Sampling and Inference*. New York. John Wiley.

Winkler, W. E. (2001). *Multi-Way Survey Stratification and Sampling, Research Report Series, Statistics* #2001-01. Statistical Research Division U.S. Bureau of the Census Washington D.C. 20233.

**Appendix A1. Note on the convergence of the fixed-point iteration**

Let

$$(A.1) \quad n_d^t = \sum_{i \in U} \pi_i^t \lambda_{di} \quad d = 1, \ldots, D$$

be the $d-$th domain sampling size of iteration $t$. They are function of the $n_d^{t-1}$ values.

We see that the minimum sample size, $n_{(v)d}^t$ $(v = 1, \ldots, V)$ to respect the $v-$th constraint of OP 2.5 is

$$(A2) \quad n_{(v)d}^t = g_{(v)}(n_d^{t-1}) = \frac{\sigma_{(v)u}^2}{V_{(v)d}^*} \left( \frac{\sigma_{(v)}^2}{n_d^{t-1}} + \sigma_{(v)u}^2 \right)^{-1} \quad (v = 1, \ldots, V; d = 1, \ldots, D).$$

where $g_{(v)}(\cdot)$ is the updating function and $V_{(v)d}^* = g1_{(v)d}^*/N_d^2$.

Indeed, we have

$$\frac{\sigma_{(v)}^2}{n_{(v)d}^t} \leq \frac{V_{(v)d}^*}{\sigma_{(v)u}^2} \left( \frac{\sigma_{(v)}^2}{n_d^{t-1}} + \sigma_{(v)u}^2 \right).$$

We have:

$$(A3) \quad n_d^t \geq n_{(v)d}^t \quad \text{for } (v = 1, \ldots, V),$$

since the sample size should ensure the respect of all of the $V$ constraints of OP 5.2.

Let $\boldsymbol{g}^t = \{ g_{(v)}(n_d^{t-1}); v = 1, \ldots, V; d = 1, \ldots, D \}$, be the system of $V \times D$ updating equations.

Let

$$(A4) \quad \boldsymbol{J}^t = diag \left\{ \frac{\partial g_{(v)}(n_d^{t-1})}{\partial n_d^{t-1}}; v = 1, \ldots, V; d = 1, \ldots, D \right\}$$

be the Jacobian of $\boldsymbol{g}^t$. We see that the

$$(A5) \quad |\boldsymbol{J}^t| \leq 1,$$

since

$$\frac{\partial g_{(v)}(n_d^{t-1})}{\partial n_d^{t-1}} = \frac{\sigma_{(v)}^2 \sigma_{(v)u}^2}{V_{(v)d}^*} \left( \frac{\sigma_{(v)}^2}{n_d^{t-1}} + \sigma_{(v)u}^2 \right)^{-2} \frac{\sigma_{(v)}^2}{(n_d^{t-1})^2} \leq$$

$$\leq \frac{\sigma_{(v)}^2 \sigma_{(v)u}^2}{V_{(v)d}^*} \left(\frac{\sigma_{(v)}^2}{n_d^{t-1}}\right)^{-2} \frac{\sigma_{(v)}^2}{(n_d^{t-1})^2}$$

$$= \frac{\frac{\sigma_{(v)}^2 \sigma_{(v)u}^2}{V_{(v)d}^*}}{\frac{\sigma_{(v)}^4}{(n_d^{t-1})^2}} \frac{\sigma_{(v)}^2}{(n_d^{t-1})^2} = \frac{\sigma_{(v)}^2 \sigma_{(v)u}^2}{V_{(v)d}^*} \frac{(n_d^{t-1})^2}{\sigma_{(v)}^4} \frac{\sigma_{(v)}^2}{(n_d^{t-1})^2} = \frac{\sigma_{(v)u}^2}{V_{(v)d}^*} < 1.$$

After having determined the $n_{(v)d}^t$ values, the final $n_d^t$ values at iteration $t$ are determined by solving OP 7.20. Result A5 shows that the relative impact on $n_{(v)d}^t$ values decrease iteration after iteration. Therefore, we can argue that the algorithm converges to a fixed-point solution.